\documentclass[prd,onecolumn,notitlepage,eqsecnum,nofootinbib,floatfix]{revtex4-1}
\usepackage{amssymb}
\usepackage{amsmath}
\usepackage{amsfonts} 
\usepackage{bm}
\usepackage{graphicx}
\usepackage{comment} 
\DeclareSymbolFont{EulerScript}{U}{eus}{m}{n}
\SetSymbolFont{EulerScript}{bold}{U}{eus}{b}{n}
\DeclareSymbolFontAlphabet\scrpt{EulerScript}

\newcommand{\gothg}{\mathfrak{g}}
 
\newcommand{\sgn}{\mbox{sign}}
\allowdisplaybreaks
\begin{document}
\title{Tidally induced multipole moments of a nonrotating black hole vanish to all post-Newtonian orders}  
\author{Eric Poisson}  
\affiliation{Department of Physics, University of Guelph, Guelph, Ontario, N1G 2W1, Canada} 
\date{November 16, 2021} 
\begin{abstract}
The tidal Love numbers of a black hole vanish, and this is often taken to imply that the hole's tidally induced multipole moments vanish also. An obstacle to establishing a link between these statements is that the multipole moments of individual bodies are not defined in general relativity, when the bodies are subjected to a mutual gravitational interaction. In a previous publication [Phys.\ Rev.\ D {\bf 103}, 064023 (2021)] I promoted the view that individual multipole moments can be defined when the mutual interaction is sufficiently weak to be described by a post-Newtonian expansion. In this view, a compact body is perceived far away as a skeletonized post-Newtonian object with a multipole structure, and the multipole moments can then be related to the body's Love numbers. I expand on this view, and demonstrate that all static, tidally induced, mass multipole moments of a nonrotating black hole vanish to all post-Newtonian orders. The proof rests on a perturbative solution to the Einstein-Maxwell equations that describes an electrically charged particle placed in the presence of a charged black hole. The gravitational attraction between particle and black hole is balanced by electrostatic repulsion, and the system is in an equilibrium state. The particle provides a tidal environment to the black hole, and the multipole moments vanish for this environment. I argue that the vanishing is robust, and applies to all slowly-varying tidal environments. The black hole's charge can be as small as desired (though not identically zero); by continuity, the multipole moments of an electrically neutral black hole will continue to vanish.
\end{abstract} 
\maketitle

\section{Introduction} 
\label{sec:intro} 

The tidal deformation of compact bodies (neutron stars and/or black holes) was revealed \cite{flanagan-hinderer:08} to play an important role in the emission of gravitational waves during a binary inspiral, and to disclose important information regarding the body's internal constitution (see Ref.~\cite{chatziioannou:20} for a recent review of these exciting developments). A measurement of the tidal deformability of a neutron star was attempted for GW170817 \cite{GW170817:17}, and it delivered an upper bound of astrophysical significance \cite{GW170817:18}; this bound was used to constrain the equation of state of nuclear matter at high densities \cite{landry-essick-reed-chatziioannou:20}.

The tidal deformability of a compact object is measured primarily by its (gravitoelectric) Love numbers $k_\ell$, which depend on the equation of state for a neutron star, and vanish for a black hole \cite{damour-nagar:09, binnington-poisson:09}. This result, that all Love numbers are zero for a black hole, has never ceased to fascinate and mystify \cite{chirenti-posada-guedes:20, hui-etal:21a, chia:21, charalambous-dubovsky-ivanov:21a}. Porto described it as a fine tuning that requires an explanation \cite{porto:16}; a hidden ladder symmetry was recently uncovered \cite{charalambous-dubovsky-ivanov:21b, hui-etal:21b} and proposed as such an explanation.  

My purpose with this paper is to clarify the link between Love numbers and observable tidal effects in a binary inspiral. It often seems to be taken for granted that the vanishing of Love numbers for a black hole implies the absence of tidal effects in the hole's equations of motion, and in the emission of gravitational waves. I believe that this conclusion should not be taken for granted; it should not be accepted without proof. Within the limitations to be specified below, I provide a proof in this paper. The conclusion turns out to be correct. 

The tendency to take the link for granted comes from Newtonian gravity, which offers no essential distinction between Love numbers and tidally induced multipole moments. To keep the discussion simple I restrict my attention to a star of mass $M$ and radius $R$ subjected to a quadrupolar ($\ell = 2$) tidal field. A calculation of the exterior Newtonian potential for this situation (see, for example, Sec.~2.5 of Ref.~\cite{poisson-will:14}) returns
\begin{equation}
U = \frac{GM}{r} - \frac{1}{2} \biggl( r^2 + 2 k_2 \frac{R^5}{r^3} \biggr)
{\cal E}_{ab} \Omega^a \Omega^b,
\label{U_vs_k} 
\end{equation}
where $r$ is the distance to the star's center of mass, $\Omega^a = x^a/r$ is the unit radial vector, $k_2$ is the star's quadrupolar Love number, and ${\cal E}_{ab}(t) := -\partial_{ab} U^{\rm ext}$ is the tidal quadrupole moment, given by two derivatives of the external potential created by the remote bodies responsible for the tidal environment (this is evaluated at the center of mass after differentiation). The potential $U$ is seen to be a superposition of two elementary solutions to Laplace's equation. The first is the growing term proportional to $r^2$, which describes the tidal field. The second is the decaying term proportional to $r^{-3}$, which describes the star's tidal response. In Newtonian gravity, decaying solutions to Laplace's equation are in a one-to-one correspondance with multipole moments of the mass distribution. A general solution to Laplace's equation that includes monopole and quadrupole terms can always be written as
\begin{equation}
U = \frac{GM}{r} - \frac{1}{2} r^2\, {\cal E}_{ab} \Omega^a \Omega^b
+ \frac{3}{2 r^3} Q_{ab} \Omega^a \Omega^b, 
\label{U_vs_Q} 
\end{equation}
where $Q_{ab}$ is the mass quadrupole moment. Comparison between Eqs.~(\ref{U_vs_k}) and (\ref{U_vs_Q}) reveals that
\begin{equation}
Q_{ab} = -\frac{2}{3} k_2 R^5 {\cal E}_{ab}
\label{Q_vs_E_short}
\end{equation}
for a tidally deformed star. In Newtonian theory, therefore, $Q_{ab}$ is proportional to $k_2$, and the Love number is essentially synonymous with the mass quadrupole moment.

The link between $k_2$ and $Q_{ab}$ is far more tenuous in general relativity. There is no obstacle to promoting Eq.~(\ref{U_vs_k}) and providing $k_2$ with a proper relativistic definition. The Newtonian potential is replaced with the metric $g_{\alpha\beta}$ of a tidally deformed body, and we obtain relations of the form \cite{damour-nagar:09, binnington-poisson:09}
\begin{equation} 
c^{-2} g_{tt} = -1 + \frac{2GM}{c^2 r}
- \frac{1}{c^2} \biggl[ r^2 (1 + \cdots)
+ 2 k_2 \frac{R^5}{r^3} (1 + \cdots) \biggr]\,
{\cal E}_{ab} \Omega^a \Omega^b,
\label{g_vs_k}
\end{equation}
where the ellipses represent relativistic corrections of order $GM/(c^2 r)$. The important point here is that in the relativistic theory, the Love numbers are defined as (gauge-invariant) parameters of the deformed metric, and nothing more. Gralla pointed out \cite{gralla:18} that the definition of Love numbers is based on a convention regarding the characterization of growing and decaying solutions in general relativity, and that different authors might choose different conventions and obtain different values. 

There is no direct analogue to Eq.~(\ref{U_vs_Q}) in general relativity, and as a consequence, there is no direct analogue to Eq.~(\ref{Q_vs_E_short}). The obstacle is the presence of growing terms in the metric. While multipole moments of a stationary and asymptotically flat spacetime can be defined rigorously \cite{geroch:70, hansen:74}, these definitions do not apply when consideration is given only to a small neighborhood of the deformed body, with a metric of the form of Eq.~(\ref{g_vs_k}); this portion of spacetime is not asymptotically flat. In other words, while the Geroch-Hansen algorithm can be exploited to calculate the multipole moments of an entire spacetime (given stationarity and asymptotic flatness), it cannot be used to compute the moments of individual objects within this spacetime. Some authors have proposed a way out \cite{pani-etal:15a, letiec-casals:21, letiec-casals-franzin:21}: subtract out the growing solution from Eq.~(\ref{g_vs_k}) and then calculate the Geroch-Hansen multipole moments. As I argue at length in Ref.~\cite{poisson:21a}, this approach is both artificial and ambiguous; it brings obscurity to an effort to elucidate the tidal dynamics of a binary system in general relativity. 

In Ref.~\cite{poisson:21a} I introduced what I consider to be a more promising approach to the definition of tidally induced multipole moments. It relies on a post-Newtonian description of the mutual gravity between the compact body and its binary companion, which is assumed to be weak. In this view, the gravitational field of the compact body becomes indistinguishable, sufficiently far away, from the field of a point particle endowed with a multipole structure. In this approximate way, the body becomes a skeletonized post-Newtonian object (I prefer to use this phrase instead of ``point particle''), and the multipole moments are properties of this object. Given this definition, a relation such as Eq.~(\ref{Q_vs_E_short}) can then be derived, implicating the post-Newtonian $Q_{ab}$ and ${\cal E}_{ab}$ and a $k_2$ computed in full general relativity. 

This approach, I believe, provides a satisfactory definition of multipole moments for strongly self-gravitating bodies, provided only that they are in a weak gravitational interaction with other masses. The grounding of the approach in post-Newtonian theory is also natural, for it is within this framework that tidal terms in the equations of motion, and in the emission of gravitational waves, are usually identified \cite{damour-nagar:10, bini-damour-faye:12, vines-flanagan:13, henry-faye-blanchet:20a}. To be clear, the description of tidal effects does not require post-Newtonian theory; they can very well, for example, be witnessed directly in numerical relativity simulations. However, the characterization of tidal effects in terms of quantities such as Love numbers and tidally induced multipole moments must be grounded in post-Newtonian theory. 

It is within this framework that I demonstrate that {\it the tidally induced multipole moments of a black hole are all zero}. With a few qualifiers inserted, what I shall show is that all static, tidally induced, mass multipole moments of a nonrotating black hole, as defined properly in terms of a skeletonized post-Newtonian object, vanish to all post-Newtonian orders. It is important to appreciate the distinction between this statement and the long-known fact that Love numbers vanish for black holes. As I have emphasized, Love numbers are properties of the deformed metric, while tidally induced multipole moments are properties of the skeletonized post-Newtonian object. To show that the vanishing of Love numbers implies the vanishing of multipole moments requires a satisfactory link between these quantities. The post-Newtonian approach provides such a link, and it provides a method of proof. 

The argument is unfolded in the next two sections. I begin in Sec.~\ref{sec:compact} with a detailed explanation of the notion that a compact body in a weak gravitational interaction with other masses can be viewed as a skeletonized post-Newtonian object. This notion gives rise to a proper definition for the tidally induced multipole moments of the compact body, and it permits a calculation of these moments. 

I continue in Sec.~\ref{sec:BHstatic} with a computation of tidally induced multipole moments for a black hole placed in a static tidal environment; these are calculated to all post-Newtonian orders. The environment that permits this is a contrived one: the black hole is given a small electric charge (as small as desired, though not identically zero) and placed in the presence of a particle with a large charge-to-mass ratio. The gravitational and electrostatic forces acting on the black hole and particle are balanced, and the system is in an equilibrium state. The particle produces a tidal field around the black hole, and mass multipole moments are computed for this field; I obtain a vanishing result for all of them. 

Should one be bothered by the contrivance? After all, an astrophysical black hole will never be found in such an arrangement. I argue that one should not be bothered. To be sure, the charged particle produces a tidal tensor ${\cal E}_{ab}$ that is highly specialized. The relation between $Q_{ab}$ and ${\cal E}_{ab}$, however, can be expected to be generic. If $Q_{ab}$ vanishes for a contrived ${\cal E}_{ab}$, it will vanish also for a naturally occurring ${\cal E}_{ab}$. To the extent that this statement is correct, the conclusion that all multipole moments vanish for a black hole will be valid for all static tidal environments.

Should one be bothered by the requirement that the black hole be charged? Again my answer is no. As I pointed out, the charge can be as small as one desires, and such a small charge will not change the properties of the black hole by much. By continuity, we can expect that if the multipole moments vanish for a very small charge, they will continue to vanish for a zero charge.

Should one be bothered by the static tidal environment? No. The idealization of a static tidal field has been exploited in most of the literature on this subject. It captures the idea that when viewed on a time scale commensurate with the black hole, the tidal environment changes slowly, thanks to the weak gravitational interaction with its companion. (An extended discussion of this point will be found in Sec.~\ref{sec:compact}.)

The calculation presented in Sec.~\ref{sec:BHstatic} requires Eq.~(\ref{W_multipole}) below as a key technical input; this gives the metric of the tidally deformed black hole. The following three sections of the paper are devoted to a derivation of this equation. These computations are long and detailed, and in order to preserve the flow of the demonstration (which is more interesting than the technical matter), I present them after the argument has been brought to a close. I begin in Sec.~\ref{sec:RN} with a review of relevant aspects of the Reissner-Nordstr\"om spacetime, which describes an electrically charged black hole. I introduce a particle with an arbitrary charge-to-mass ratio in Sec.~\ref{sec:pert-general}, and calculate the gravitational and electromagnetic perturbations that it produces; in this general context the particle and black hole are kept in place by means of massless strings. The perturbation is a solution to the linearized Einstein-Maxwell equations, and the perturbed spacetime is an approximation to an exact solution that describes the superposition of two Reissner-Nordstr\"om objects \cite{manko:07, manko-ruiz-sanchezmondragon:09}; in view of its simplicity compared with the exact spacetime, I prefer to use the perturbative solution. In Sec.~\ref{sec:pert-equil}, I specialize the solution to the specific charge-to-mass ratio that produces balanced gravitational and electrostatic forces. This situation was previously examined in Ref.~\cite{bini-geralico-ruffini:07}; my solution is expressed in a different (and more convenient) gauge. An exact version of this spacetime was obtained by Alekseev and Belinski \cite{alekseev-belinski:07}; again the linearized solution is simpler to deal with. After all these developments, I arrive at Eq.~(\ref{W_multipole}). 

Some calculations are relegated to appendices. I construct monopole solutions to the perturbed Einstein-Maxwell equations in Appendix~\ref{sec:monopole}; these are required in Sec.~\ref{sec:pert-general}. I examine the dipole piece of the perturbation in Appendix~\ref{sec:dipole}, and introduce a coordinate transformation that eliminates the solution's dipole moment at infinity. 

\section{Compact body as skeletonized post-Newtonian object}
\label{sec:compact}

We imagine a compact body (a neutron star, a black hole, or even something more exotic) placed in the presence of other masses, such that the entire system forms a bound gravitating system. We imagine that the mutual gravitational interaction between these bodies is sufficiently weak that it can be adequately described by a post-Newtonian expansion. On the other hand, gravity is strong in the immediate neighborhood of the compact body, and it must be given a fully relativistic description. For conceptual simplicity we assume that the internal gravity of each remote object is weak, so that it is also well described by a post-Newtonian expansion; there is no obstacle to relaxing this assumption, and allowing the internal gravity of all bodies to be strong.

\begin{figure}
\includegraphics[width=0.7\linewidth]{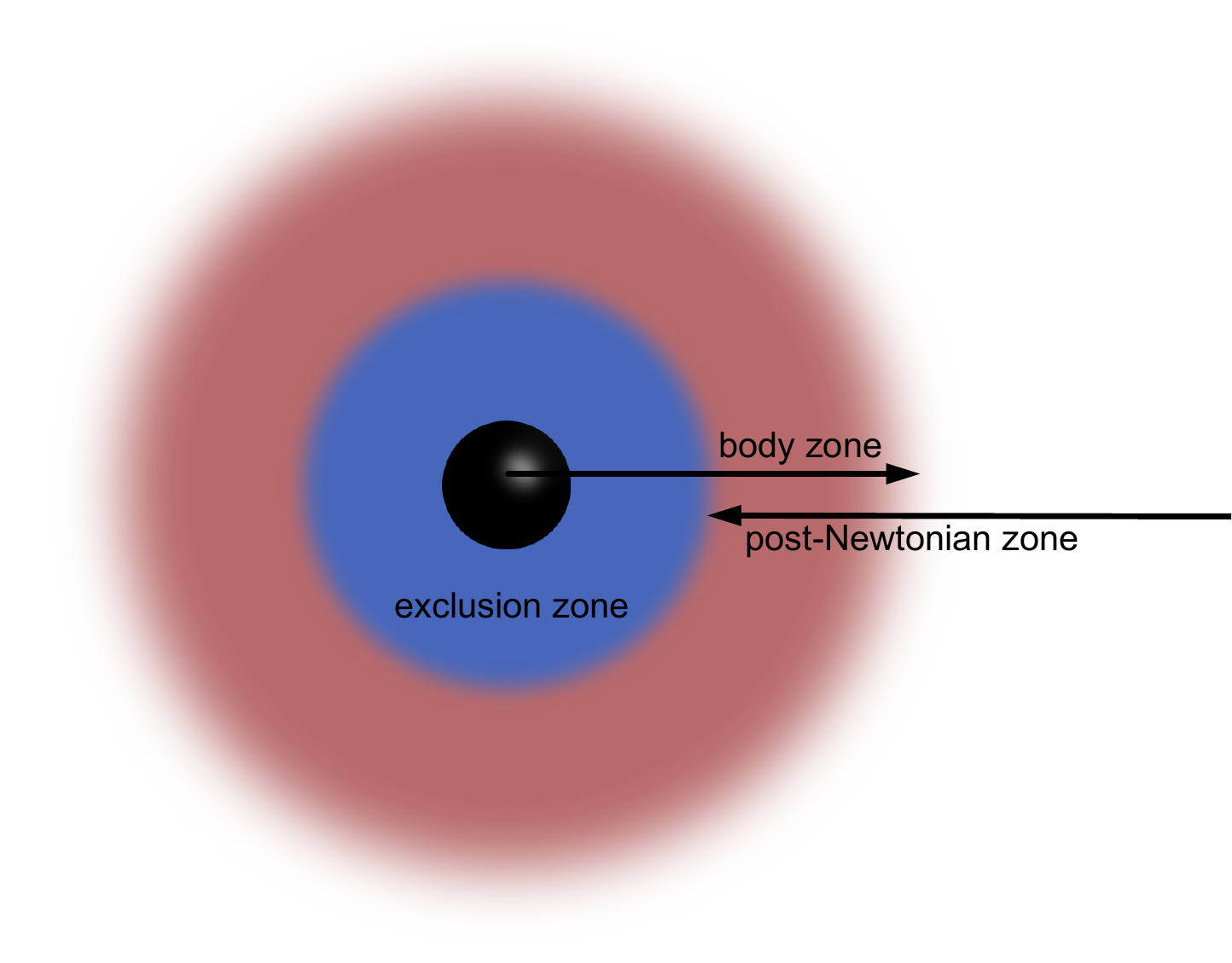}
\caption{Spacetime partitioned into zones. The black hole is shown at the center, in black. The post-Newtonian zone, where gravity is weak, is shown in white. The exclusion zone, where gravity is strong, is shown in blue. The body zone, where gravity is described by a tidal deformation of the original black-hole spacetime, is shown in red.} 
\label{fig:fig1} 
\end{figure} 

We wish to describe the weak mutual interaction between compact body and remote objects, while taking into account the fact the body's own gravity is strong. For this purpose we partition spacetime near the compact body into zones --- see Fig.~\ref{fig:fig1}. The first is a {\it post-Newtonian zone}, in which gravity is weak, and in which the metric can be given a post-Newtonian expansion. The second is an {\it exclusion zone}, which includes and surrounds the compact body, in which gravity is strong, and in which the metric must be calculated in full general relativity. (A third zone, the {\it body zone}, will be introduced below.) The post-Newtonian zone can be further partitioned into {\it near} and {\it wave} zones; in the near zone gravity is slowly varying, with a characteristic velocity that is small compared with the speed of light, while it is rapidly varying in the wave zone, with a characteristic velocity equal to the speed of light. Our considerations below will be restricted to the near zone. 

In the post-Newtonian zone, the Einstein field equations are cast in the form first introduced by Landau and Lifshitz \cite{landau-lifshitz:b2} (see Chapter 6 of Ref.~\cite{poisson-will:14} for an elaboration of their formalism). The primary variable is the ``gothic inverse metric'' $\gothg^{\alpha\beta} := \sqrt{-g}\, g^{\alpha\beta}$, where $g^{\alpha\beta}$ is the inverse of the actual metric $g_{\alpha\beta}$, and $g := \mbox{det}[g_{\alpha\beta}]$. We impose the harmonic coordinate conditions
\begin{equation} 
\partial_\beta \gothg^{\alpha\beta} = 0,
\label{harmonic_gothg} 
\end{equation}
which are formally equivalent to the statement that the four spacetime coordinates $X^\mu = (t, X^a)$ satisfy the wave equation $g^{\alpha\beta} \nabla_\alpha \nabla_\beta X^\mu = 0$ in a spacetime with metric $g_{\alpha\beta}$. We assume that the coordinates are Lorentzian, in the sense that $\gothg^{\alpha\beta}$ approaches the Minkowski metric at infinity. The field equations become
\begin{equation}
\nabla^2 \gothg^{\alpha\beta} = \frac{1}{c^2} \partial_{tt} \gothg^{\alpha\beta} 
+ \frac{16\pi G}{c^4} \tau^{\alpha\beta},
\label{EFE_LL} 
\end{equation}
where $\nabla^2$ is the usual flat-space Laplacian, $\partial_t$ denotes partial differentiation with respect to time, and $\tau^{\alpha\beta}$ is an effective energy-momentum pseudotensor which contains contributions from the matter and the gravitational field.  

Equation (\ref{EFE_LL}), accompanied by Eq.~(\ref{harmonic_gothg}), is an exact formulation of the Einstein field equations. Building on this, the post-Newtonian approximation introduces a simultaneous expansion of $\gothg^{\alpha\beta}$ in powers of $G$ and $c^{-2}$, which must keep step when the system is gravitationally bound. Equation (\ref{EFE_LL}) is then iterated as many times as required to achieve a desired degree of accuracy. When the scheme is implemented in the near zone, the zeroth iteration sets the metric to its Minkowski value on the right-hand side of the equation, and the resulting Poisson equation is integrated. In the first iteration the solution is inserted on the right-hand side, and the new Poisson equation is integrated. And so on. The key point is that during each iteration of the field equations, the source term on the right-hand side is determined from the previous iteration.

At each iteration the integration of Poisson's equation is carried out in the near zone only. The domain of integration therefore excludes the wave zone and the exclusion zone. These, however, must inform the solution, and the relevant information is fed through solutions to the homogeneous version of Eq.~(\ref{EFE_LL}), Laplace's equation $\nabla^2 \gothg^{\alpha\beta} = 0$. From the wave zone we get growing solutions of the form $\bar{r}^\ell Y_\ell^m(\theta,\phi)$, where $(\bar{r},\theta,\phi)$ are spherical polar coordinates obtained in the usual way from the Cartesian system $X^a$, and where $Y_\ell^m$ are spherical harmonics. These solutions reflect a choice of outgoing-wave boundary condition at infinity, and they give rise to radiation-reaction forces acting on the system. From the exclusion zone we get decaying solutions of the form $\bar{s}^{-\ell-1} Y_\ell^m(\theta,\phi)$, where $\bar{s} := |\bm{X} - \bm{X}_{\rm body}|$ is the distance to the compact body. Such a term is formally identical to a contribution to $\gothg^{\alpha\beta}$ that would come from a multipole of order $\ell$ situated at $X^a = X^a_{\rm body}(t)$. In this way, the compact body manifests itself, in the post-Newtonian zone, as a skeletonized object with a collection of multipole moments. Two types of moments occur in the metric: mass multipole moments, which appear primarily in $\gothg^{tt}$, and current multipole moments, which appear primarily in $\gothg^{ta}$. 

The subsumption of the compact body into a skeletonized post-Newtonian object with a multipole structure\footnote{A question might be raised about the gauge invariance of the multipole moments. I claim that they are indeed gauge invariant, in the following sense. Note first that the harmonic coordinates of the post-Newtonian metric are uniquely specified (assuming regularity at the origin, an outgoing-wave boundary condition at infinity, and up to a global rotation and translation). Given that the harmonic gauge is completely fixed, quantities that appear in the post-Newtonian metric are then necessarily gauge invariant, and this is true of the multipole moments of the skeletonized body. The multipole moments, however, are not directly observable. But they can be mapped to an observable \cite{damour-nagar:10, bini-damour-faye:12, vines-flanagan:13, henry-faye-blanchet:20a}: they appear in the expression of the gravitational waves measured at infinity, which are directly observable.} is an artifact of the post-Newtonian approximation and the need to truncate the post-Newtonian zone when the body's gravity becomes too strong. The actual body, of course, is not pointlike. To determine the multipole moments it is necessary to obtain an alternative description of the metric, in a region of spacetime --- the {\it body zone} --- that includes the body and extends beyond the exclusion zone, in a treatment that is now grounded in full general relativity. In this description the compact body is taken to be weakly deformed by the tidal field created by the remote objects; the body would be spherical in isolation. The tidal deformation is measured in terms of Love numbers, which appear as parameters in the body-zone metric, and which are computed by ensuring that the body's exterior and interior metrics join smoothly at the surface. The body's local metric is then matched to the post-Newtonian metric in the overlap between body and post-Newtonian zones, and the mass and current multipole moments are determined in terms of the Love numbers. It is important to understand that the Love numbers are a property of the body's local metric, which is obtained in full general relativity, and that the multipole moments are a property of the skeletonized post-Newtonian object. It is the matching of these two disparate descriptions of the same gravitational field that produces a relationship between Love numbers and tidally induced multipole moments.

The matching exercise described in the preceding paragraph is difficult to perform in practice. First, a post-Newtonian metric describing a skeletonized body with an arbitrary multipole structure in a bound interaction with remote objects must be constructed to a sufficient degree of accuracy. Second, this metric must be transformed from the original barycentric frame of the entire system (taken to be at rest) to the body's own rest frame. Third, the metric of a tidally deformed body must be computed in full general relativity, with all the required ingredients that ensure a proper match with the post-Newtonian metric. And fourth, the matching must be carried out. Thus far, this procedure has been completed to the first post-Newtonian order only \cite{poisson:21a}. It produced the relation
\begin{equation}
Q_{ab} = -\frac{2}{3} k_2 R^5\, {\cal E}_{ab}
- \frac{2}{3} p_2 \frac{R^8}{GM}\, {\cal E}_{c\langle a} {\cal E}^c_{\ b\rangle}
- \frac{2}{3} \ddot{k}_2 \frac{R^8}{GM}\, \ddot{\cal E}_{ab}
+ \mbox{post-Newtonian corrections},
\label{Q_vs_E} 
\end{equation}
where $Q_{ab}$ is the mass quadrupole moment of the skeletonized compact body, $M$ is the body's mass, $R$ its radius, $k_2$, $p_2$, $\ddot{k}_2$ are Love numbers, and where
\begin{equation}
{\cal E}_{ab}(t) = -\partial_{ab} U^{\rm ext}
+ \mbox{post-Newtonian corrections},
\end{equation}
with $U^{\rm ext}$ denoting the Newtonian potential created by the remote masses, which is differentiated twice with respect to the spatial coordinates and then evaluated at the origin of the body's rest frame. Overdots on ${\cal E}_{ab}$ indicate differentiation with respect to time, and angular brackets around spatial indices are an instruction to symmetrize the indices and remove the trace; all tensors in Eq.~(\ref{Q_vs_E}) are symmetric and trace-free. 

Equation (\ref{Q_vs_E}) generalizes Eq.~(\ref{Q_vs_E_short}), which neglected nonlinear and time-dependent terms, and which did not incorporate post-Newtonian corrections. The interpretation of Eq.~(\ref{Q_vs_E}), however, is vastly different. As was observed in Sec.~\ref{sec:intro}, the link between $Q_{ab}$ and the Love numbers is immediate in Newtonian gravity. This is not so here. As was emphasized in the preceding paragraphs, the quadrupole moment is a property of the compact body in its post-Newtonian manifestation as a skeletonized object; its definition is rooted in the post-Newtonian framework and its reliance on harmonic coordinates. The Love numbers, on the other hand, are a property of the body's local metric, and are computed in full general relativity. The relation between multipole moments and Love numbers is provided by the matching procedure. 

Equation (\ref{Q_vs_E}) applies to a material body such as a neutron star. It might be tempting to apply the relation to a black hole, for which $k_2 = p_2 = \ddot{k}_2 = 0$, and conclude that $Q_{ab} = 0$. This is not justified. The reason is that $R$ no longer provides an independent characterization of the compact body; it is now firmly equal to $2GM/c^2$. In the calculations described previously, factors of $R$ did not matter in the determination of post-Newtonian order. This changes when $R = 2GM/c^2$. For example, the first term on the right-hand side of Eq.~(\ref{Q_vs_E}) is promoted from Newtonian order to the fifth post-Newtonian order. The equation, however, was derived on the basis of a post-Newtonian expansion carried out to the first order only, and it simply cannot be applied to a black hole. A proper accounting of the mass quadrupole moment of a tidally deformed black hole requires an expansion pushed at least to the fifth post-Newtonian order. 

\section{Black hole in a static binary system}
\label{sec:BHstatic}

The current state of development of post-Newtonian theory does not allow us to perform these calculations --- metric construction for a skeletonized body with an arbitrary multipole structure, transformation to the body's rest frame, and matching with the body's local metric --- to the fifth post-Newtonian order. In an effort to make progress we shall formulate a special case that allows us to perform the calculations to all post-Newtonian orders. The situation, to be sure, is a contrived one that will never be realized in nature. This is of no concern. We hold firm onto the principle that a black hole's response to a tidal field, as described by a relation between $Q_{ab}$ and ${\cal E}_{ab}$ of the kind shown in Eq.~(\ref{Q_vs_E}), will be the same irrespective of the precise specification of ${\cal E}_{ab}$. An astrophysical tidal environment will produce a ${\cal E}_{ab}$ that is very different from the ${\cal E}_{ab}$ of a contrived environment. The relation between $Q_{ab}$ and ${\cal E}_{ab}$, however, will be the same. In particular, if $Q_{ab}$ is found to vanish for the contrived ${\cal E}_{ab}$, then it will vanish also for the realistic ${\cal E}_{ab}$.

From this point onward we adopt relativistic units, and set $G = c = 1$. 

We consider a static binary system consisting of a Reissner-Nordstr\"om (RN) black hole of mass $M$ and charge $Q$ and a point particle of mass $m$ and charge $q$. We place the particle at a distance $r = r_0$ from the black hole (in the usual RN coordinates), and we tune its charge-to-mass ratio so that
\begin{equation}
m \bigl(M - Q^2/r_0 \bigr) = q Q \bigl( 1 - 2M/r_0 + Q^2/r_0^2 \bigr)^{1/2}.  
\label{ratio} 
\end{equation}
This condition ensures --- see Sec.~\ref{sec:RN} for details --- that the particle's acceleration in the RN spacetime vanishes: the gravitational attraction between black hole and particle is precisely balanced by electrostatic repulsion. The binary system is in an equilibrium state\footnote{In a preliminary version of the work I attempted to keep the black hole and particle apart with the help of a strut or strings, instead of using charges. This, unfortunately, produced an unwieldy spacetime. In the case of a strut, the black hole was deformed in part by the particle's tidal field, but also by the strut, and this made the interpretation of results difficult. In the case of strings, the spacetime was not asymptotically flat, the harmonic coordinates were complicated, and again the interpretation of results was difficult.}; the spacetime is static, and also axisymmetric with respect to the axis that links black hole and particle. The equilibrium is unstable, but this need not concern us; we can imagine that a control system maintains the equilibrium over a sufficiently long time.   

We let the particle create a perturbation of the RN spacetime. The perturbation is in part gravitational and in part electromagnetic, and when observed near the black hole, it describes a tidal deformation. When viewed far away from the black hole, where post-Newtonian theory provides a valid description of the metric, the perturbation describes the multipole structure of the skeletonized black hole. We wish to determine this multipole structure.

The static and axially symmetric nature of the spacetime facilitates the integration of the Einstein-Maxwell equations for the perturbation. We construct the solution in Secs.~\ref{sec:RN}, \ref{sec:pert-general}, and \ref{sec:pert-equil}, and express it in closed form, to all post-Newtonian orders. The solution can easily be transformed to harmonic coordinates, a requirement for the proper identification of a post-Newtonian multipole structure.
(The black hole is situated at the spatial origin of the harmonic coordinates. There is no need here to introduce two frames of reference, one for the global post-Newtonian metric and one for the black hole, to transform between these frames, and to perform a matching.)  By carrying out a post-Newtonian expansion of the metric, we identify all terms that behave as $\bar{r}^{-\ell-1} P_\ell(\cos\theta)$ in $\gothg^{\alpha\beta}$, and read off the multipole moments of the skeletonized black hole.

It is sufficient to examine the multipole expansion of $W := -\gothg^{tt}$ when $r < r_0$, which is given by 
\begin{equation}
W(\bar{r},\theta) = W_{\rm RN} + \sum_{\ell=1}^\infty W_\ell(\bar{r})\, P_\ell(\cos\theta),
\label{W_multipole} 
\end{equation}
where
\begin{equation} 
W_{\rm RN} := \frac{(\xi+\mu)^4}{\xi^2 (\xi^2-1)}
\end{equation}
is the Reissner-Nordstr\"om piece, and 
\begin{equation}
W_\ell = -\frac{4(2\ell+1)}{\ell(\ell+1)} \frac{m}{L} (\xi_0^2-1)^{1/2} Q'_\ell(\xi_0)\,
\frac{(\xi+\mu)^3}{\xi^2} P'_\ell(\xi)
\label{Well} 
\end{equation}
makes up the perturbation. Here $\xi := \bar{r}/L$, $\xi_0 := \bar{r}_0/L$, $\mu := M/L$, and $L := (M^2-Q^2)^{1/2}$; $P_\ell(\xi)$ and $Q_\ell(\xi)$ are Legendre functions defined in the interval $1 \leq \xi <\infty$, and a prime indicates differentiation with respect to the argument. The relation between the harmonic radial coordinate $\bar{r}$ and the original RN radius $r$ is provided by $\bar{r} = r-M$. The solution of Eq.~(\ref{Well}) is regular at the event horizon, which is situated at $\xi = 1$ (or $\bar{r} = L$, or $r = M + L$). For a material body the solution would also feature a term proportional to $Q'_\ell(\xi)$, which diverges at $\xi = 1$; it would come with a numerical coefficient proportional to the object's Love number $k_\ell$. The solution of Eq.~(\ref{Well}), therefore, reflects the property that all tidal Love numbers vanish for a black hole. 

For $\ell \geq 2$, $W_\ell$ possesses all the elements that describe a tidally deformed black hole. (The dipole contribution is examined separately in Appendix~\ref{sec:dipole}.) When $\xi_0$ is large, which occurs when the particle is far away from the black hole, we have that
\begin{equation}
(\xi_0^2-1)^{1/2} Q_\ell'(\xi_0) \sim -\frac{(\ell+1)!}{(2\ell+1)!!}\, \xi_0^{-(\ell+1)}; 
\end{equation}
this displays the scaling expected of a tidal field of multipole order $\ell$. The remaining terms in an expansion of the left-hand side in powers of $\xi_0^{-1}$ represent post-Newtonian corrections to the leading behavior. When $\xi$ is large (but still smaller than $\xi_0$), which occurs when the perturbation is examined far away from the black hole, we have that
\begin{equation}
\frac{(\xi + \mu)^3}{\xi^2} P'_\ell(\xi) \sim \frac{(2\ell-1)!!}{(\ell-1)!}\, \xi^\ell; 
\label{tidal_structure} 
\end{equation}
this grows as $\xi^\ell$, again as expected of a tidal field of multipole order $\ell$. The remaining terms in an expansion of the left-hand side in powers of $\xi^{-1}$ represent post-Newtonian corrections.

And now we come to the crux of the matter. The perturbation of Eq.~(\ref{Well}) could in principle include a term that represents a tidally induced $\ell$-pole moment for the black hole; this term would scale as $\xi^{-\ell-1}$. Such a term does not exist. If we examine the left-hand-side of Eq.~(\ref{tidal_structure}), we see that it expands into a terminating polynomial, with the schematic structure
\begin{equation}
\xi^\ell + \xi^{\ell-1} + \cdots + \xi^{n},
\end{equation}
in which we omit all numerical coefficients. The smallest power is $n = -1$ when $\ell$ is even, and $n=-2$ when $\ell$ is odd. This cannot be equal to $-(\ell+1)$ when $\ell \geq 2$, as would be required for a nonvanishing tidal response. (There is such a term when $\ell = 1$, but it does not describe a genuine response to an applied tidal field; this is clarified in Appendix~\ref{sec:dipole}.) We conclude that {\it all static, tidally induced, mass multipole moments of a nonrotating black hole}, as defined in terms of a skeletonized post-Newtonian object, {\it vanish to all post-Newtonian orders}.

We insist that this conclusion is distinct from the statement that Love numbers vanish for a black hole. As was pointed out previously, Love numbers are a property of the perturbed metric, and they cannot be related directly to tidally induced multipole moments. Such a relation requires a definition of multipole moments for an individual body, which is provided here by the post-Newtonian description of the gravitational field, in which the compact body manifests itself as a skeletonized object. In principle, the determination of the multipole moments should come from a careful match between the local metric of the compact object, computed in full general relativity, and the post-Newtonian metric, properly transformed to the rest frame of the compact object. The calculation presented here bypasses most of those steps: a single frame of reference is required, and the global metric is computed in full general relativity as a perturbation of the Reissner-Nordstr\"om spacetime. The post-Newtonian expansion is introduced at the very end of the exercise, and it reveals the complete absence of a multipole structure for the skeletonized black hole. 

The tidal environment considered here is precisely static. As a result, the expected relationship between $Q_{ab}$ and ${\cal E}_{ab}$ could hope to reproduce the first term in Eq.~(\ref{Q_vs_E}), but it cannot capture the third term involving time derivatives of the tidal field. The tidal environment is also treated as a linearized perturbation of the original black-hole spacetime. As a result, the expected relationship cannot account for the second term in Eq.~(\ref{Q_vs_E}), which is quadratic in the tidal field. Within these limitations, we have a proof that in the static and linearized regime, all multipole moments vanish to all post-Newtonian orders.

The proof requires the black hole to have a charge $Q$, so that the tidal environment can be manufactured by a charged particle maintained in a state of equilibrium by balanced gravitational and electrostatic forces. The charge, however, can be as small as one desires, though not identically zero. To see this, let us rewrite Eq.~(\ref{ratio}) in the form
\begin{equation}
\frac{Q}{M} = \frac{m}{q} \frac{1 - Q^2/(M r_0)}{(1 - 2M/r_0 + Q^2/r_0^2)^{1/2}}.
\end{equation}
The second factor on the right is of order unity when the particle is not too close to the black hole, and the equilibrium condition states that the charge-to-mass ratio of the black hole must be the inverse of the particle's own ratio. By choosing a particle with $q/m \gg 1$ we can ensure that $Q/M \ll 1$. So while the black hole's charge is an essential device required in the setup of a static tidal environment, it can be chosen so small as to be utterly irrelevant in every other aspect of the problem. 

\section{Reissner-Nordstr\"om spacetime}
\label{sec:RN}

In this and the remaining sections of the paper we provide a derivation of Eq.~(\ref{W_multipole}). We begin with a review of the Reissner-Nordstr\"om (RN) spacetime, with a focus on those aspects that are germane to the discussion. 

\subsection{Metric and vector potential} 

The RN spacetime is an exact solution to the Einstein-Maxwell equations that describes a spherical black hole of mass $M$ and electric charge $Q$. The metric is written as
\begin{equation}
ds^2 = -f\, dt^2 + f^{-1}\, dr^2 + r^2(d\theta^2 + \sin^2\theta\, d\phi^2), 
\label{metric_RN} 
\end{equation}
where
\begin{equation}
f := 1 - \frac{2M}{r} + \frac{Q^2}{r^2}.
\end{equation}
The electromagnetic vector potential is given by 
\begin{equation}
A_\alpha = -\frac{Q}{r}\, \partial_\alpha t.
\label{vecpot_RN} 
\end{equation}
The electromagnetic field tensor is then $F_{\alpha\beta} = \nabla_\alpha A_\beta - \nabla_\beta A_\alpha$. For a static observer, it describes a radial electric field that behaves as $Q/r^2$.  

\subsection{Charged particle} 

We introduce, at position $r = r_0$ and $\theta = 0$ in the RN spacetime, a point particle of mass $m$ and electric charge $q$; we take $m \ll M$ and $q \ll Q$. The particle's world line is described by the parametric equations $x^\alpha = Z^\alpha(\tau)$, where $\tau$ is proper time, and its velocity vector is $u^\alpha := dZ^\alpha/d\tau$; the only nonvanishing component is $u^t = f_0^{-1/2}$, where $f_0 := 1-2M/r_0 + Q^2/r_0^2$.

The particle comes with an energy-momentum tensor
\begin{equation}
T^{\alpha\beta} = m \int u^\alpha u^\beta\, \delta(x, Z)\, d\tau,
\label{T-part} 
\end{equation}
in which $\delta(x,z) = \delta(x-Z)/\sqrt{-g}$ is a scalarized Dirac distribution; $\delta(x-Z)$ is the usual four-dimensional delta function and $g$ is the metric determinant. The only nonvanishing component is 
\begin{equation}
T^t_{\ t} = -m \frac{\sqrt{f_0}}{r_0^2}\, \delta(r-r_0) \delta(\cos\theta-1) \delta(\phi - \phi_0),
\end{equation}
where $\phi_0$ is the particle's (arbitrarily assigned) azimuthal position in the RN spacetime.

The particle also comes with a current density
\begin{equation}
j^\alpha = q \int u^\alpha\, \delta(x,Z)\, d\tau, 
\end{equation}
with
\begin{equation}
j^t = \frac{q}{r_0^2}\, \delta(r-r_0) \delta(\cos\theta-1) \delta(\phi - \phi_0)
\end{equation}
as its only nonvanishing component.

The particle's Killing energy, defined by $E := -(m u_\alpha + q A_\alpha) t^\alpha$, where $t^\alpha = [1,0,0,0]$ is the spacetime's timelike Killing vector, evaluates to 
\begin{equation} 
E = m \sqrt{f_0} + \frac{qQ}{r_0}.  
\label{killing_energy} 
\end{equation}

The force required of an external agent to keep the particle in place is
\begin{equation} 
F^\alpha = m u^\beta \nabla_\beta u^\alpha - q F^\alpha_{\ \beta} u^\beta,
\end{equation} 
and its only nonvanishing component is $F^r$. Its covariant magnitude is $F := \pm (g_{\alpha\beta} F^\alpha F^\beta)^{1/2}$, with the sign chosen so that $\mbox{sign}(F) = \mbox{sign}(F^r)$. We find that 
\begin{equation}
F = \frac{1}{r_0^2} \biggl[ \frac{m(M - Q^2/r_0)}{\sqrt{f_0}} - qQ \biggr]. 
\label{force} 
\end{equation}
In Sec.~\ref{sec:pert-equil} we shall choose $q/m$ so that $F$ vanishes: the gravitational attraction between particle and black hole shall be balanced by electrostatic repulsion. Throughout Sec.~\ref{sec:pert-general}, however, we shall keep $q/m$ general and have $F \neq 0$. 

\subsection{Harmonic coordinates} 

Each one of the scalar fields
\begin{subequations}
\label{harm_coordinates}
\begin{align}
X^0 &:= t, \\
X^1 &:= (r-M) \sin\theta\cos\phi, \\
X^2 &:= (r-M) \sin\theta\sin\phi, \\
X^3 &:= (r-M) \cos\theta 
\end{align}
\end{subequations}
satisfies the wave equation
\begin{equation} 
g^{\alpha\beta} \nabla_\alpha \nabla_\beta\, X^\mu = 0
\label{wave-eqn} 
\end{equation}
in the RN spacetime. The collection therefore makes up a set of harmonic coordinates.

In the coordinates $X^\mu = (t, X^a)$ we have that the metric and its inverse are given by  
\begin{subequations}
\begin{align}
g_{tt} &= -f, \\
g_{ab} &= f^{-1}\, \Omega_a \Omega_b
+ (1 + M/\bar{r})^2 \bigl( \delta_{ab} - \Omega_a \Omega_b \bigr), \\
g^{tt} &= -f^{-1}, \\
g^{ab} &= f\, \Omega^a \Omega^b
+ (1 + M/\bar{r})^{-2} \bigl( \delta^{ab} - \Omega^a \Omega^b \bigr),
\end{align}
\end{subequations}
where $\bar{r} := r - M$,
\begin{equation}
f = \frac{1 - L^2/\bar{r}^2}{(1+M/\bar{r})^2}
\end{equation}
with $L^2 := M^2 - Q^2$, and $\Omega^a := X^a/\bar{r}$, $\Omega_a := \delta_{ab} \Omega^a$. From this it follows that
\begin{equation}
\sqrt{-g} = (1 + M/\bar{r})^2,
\end{equation}
and the components of $\gothg^{\alpha\beta} := \sqrt{-g}\, g^{\alpha\beta}$ are
\begin{subequations}
\label{gothg_RN}
\begin{align}
\gothg^{tt} &= -\frac{(1+M/\bar{r})^4}{1-L^2/\bar{r}^2}, \\
\gothg^{ab} &= \delta^{ab} - \frac{L^2}{\bar{r}^2}\, \Omega^a \Omega^b.
\end{align}
\end{subequations}
The gothic inverse metric admits a post-Newtonian expansion in powers of $M/\bar{r}$ and $L/\bar{r}$. In this expansion, $\gothg^{tt} + 1$ begins at the Newtonian order with a term proportional to $M/\bar{r}$, while $\gothg^{ab} - \delta^{ab}$ begins and ends at the second post-Newtonian order with a term proportional to $L^2/\bar{r}^2$. (Recall that by convention, the  counting of post-Newtonian orders is different for spatial and temporal components of the metric.)    

\section{Perturbed spacetime: General case} 
\label{sec:pert-general}

In this section we incorporate the particle into the description of the gravitational and electromagnetic fields. Taking into account our assumption that $m \ll M$ and $q \ll Q$, we compute the small perturbations to the metric and vector potential created by the particle; all equations shall be linearized with respect to these perturbations. Our calculations generalize those of Ref.~\cite{bini-geralico-ruffini:07}, in which the condition of balanced gravitational and electrostatic forces was enforced. Here our charge-to-mass ratio is arbitrary and $F \neq 0$; we refer to this as the general case. The special case $F = 0$ will be recovered in Sec.~\ref{sec:pert-equil}. 

\subsection{Perturbation equations} 
\label{subsec:decoupling} 

We wish to calculate the perturbations $\delta g_{\alpha\beta}$, $\delta A_\alpha$ created by a particle of mass $m$ and charge $q$ at position $r = r_0$, $\theta = 0$; we work in the original $(t,r,\theta,\phi)$ coordinates. The perturbing energy-momentum tensor $\delta T^{\alpha\beta}$ includes contributions from the particle --- see Eq.~(\ref{T-part}) --- and from the perturbation $\delta F_{\alpha\beta}$ of the electromagnetic field. Because it satisfies $\delta T^r_{\ r} + \delta T^\theta_{\ \theta} = 0$, the perturbed metric can be cast in the form 
\begin{equation} 
ds^2 = -e^{-2U} f\, dt^2 + e^{2(U+\gamma)} \bigl( f^{-1}\, dr^2
+ r^2\, d\theta^2 \bigr) + e^{2U} r^2\sin^2\theta\, d\phi^2, 
\label{metric} 
\end{equation} 
in which $U(r,\theta)$ and $\gamma(r,\theta)$ are the perturbations; these are taken to be small, so that $e^{-2U} = 1 - 2U$ and so on. In the terminology of LaHaye and Poisson \cite{lahaye-poisson:21}, the metric perturbation belongs to the Weyl class, a special case of the Weyl gauge introduced in their paper. 

The vector potential is written as 
\begin{equation} 
A_\alpha = -\biggl( \frac{Q}{r} + \Phi \biggr) \partial_\alpha t, 
\label{vector_potential} 
\end{equation} 
where $\Phi(r,\theta)$ is the perturbation.

The Einstein-Maxwell equations, linearized about the RN solution, decouple into a system of equations for $U$ and $\Phi$, 
\begin{subequations}
\label{UP_eqns} 
\begin{align}
r^2 f\, \partial_{rr} \Phi + 2rf\, \partial_r \Phi + \partial_{\theta\theta} \Phi
+ \frac{\cos\theta}{\sin\theta}\, \partial_\theta \Phi - 2Q f\, \partial_r U
&= -2q f_0\, \delta(r-r_0) \delta(\cos\theta-1),  \\
r^2 f\, \partial_{rr} U + 2(r-M)\, \partial_r U + \partial_{\theta\theta} U
+ \frac{\cos\theta}{\sin\theta}\, \partial_\theta U + \frac{2Q^2}{r^2}\, U 
- 2 Q\, \partial_r \Phi &= -2m \sqrt{f_0}\, \delta(r-r_0) \delta(\cos\theta-1),
\end{align}
\end{subequations} 
and equations for $\gamma$, 
\begin{subequations}
\label{g_eqns} 
\begin{align} 
\frac{(r-M)^2-L^2\cos^2\theta}{2\sin^2\theta}\, \partial_r \gamma
&= -(r-M)(M-Q^2/r)\, \partial_r U
- \frac{(M-Q^2/r)\cos\theta}{\sin\theta}\, \partial_\theta U
- \frac{Q^2 (r-M)}{r^2}\, U 
\nonumber \\ & \quad \mbox{}
+ Q(r-M)\, \partial_r \Phi
+ \frac{Q\cos\theta}{\sin\theta}\, \partial_\theta \Phi, \\
\frac{(r-M)^2-L^2\cos^2\theta}{2\sin^2\theta}\, \partial_\theta \gamma
&= \frac{r^2 f(M-Q^2/r)\cos\theta}{\sin\theta}\, \partial_r U
- (r-M)(M-Q^2/r)\, \partial_\theta U
+ \frac{Q^2 f \cos\theta}{\sin\theta}\, U
\nonumber \\ & \quad \mbox{}
- \frac{Q r^2 f\cos\theta}{\sin\theta}\, \partial_r \Phi
+ Q(r-M)\, \partial_\theta \Phi.  
\end{align}
\end{subequations} 
Equations (\ref{UP_eqns}) were averaged over $\phi$ to eliminate the irrelevant factor $\delta(\phi-\phi_0)$ in the source terms. 

Equations (\ref{UP_eqns}) are decoupled by writing 
\begin{equation} 
\Phi = (M - Q^2/r) (\sqrt{f} A) + Q f B, \qquad 
U = (M-Q^2/r) B + Q (\sqrt{f} A), 
\label{decoupling} 
\end{equation} 
where $A(r,\theta)$ and $B(r,\theta)$ are new perturbation variables. They satisfy 
\begin{subequations} 
\label{eqAB} 
\begin{align} 
\partial_r (r^2 f \partial_r A) + \frac{1}{\sin\theta} \partial_\theta (\sin\theta\, \partial_\theta A) 
- \frac{L^2}{r^2 f} A &= -2\Gamma_A \delta(r-r_0) \delta(\cos\theta-1),  
\label{eqA} \\ 
\partial_r (r^2 f \partial_r B) + \frac{1}{\sin\theta} \partial_\theta (\sin\theta\, \partial_\theta B) 
&= -2\Gamma_B \delta(r-r_0) \delta(\cos\theta-1),  
\label{eqB}
\end{align} 
\end{subequations}
where 
\begin{subequations} 
\label{Gdef} 
\begin{align} 
\Gamma_A &:= \frac{1}{L^2} \bigl[ q(M-Q^2/r_0) \sqrt{f_0} - m Q f_0 \bigr], \\ 
\Gamma_B &:= \frac{1}{L^2} \bigl[ m(M-Q^2/r_0) \sqrt{f_0} - q Q f_0 \bigr]. 
\end{align} 
\end{subequations} 
The factor of $f^{1/2}$ placed in front of $A$ in Eq.~(\ref{decoupling}) is not required for the decoupling; it is introduced so that the differential operators acting on $A$ and $B$ come out similar.  

\subsection{Mode solution to the perturbation equations} 
\label{subsec:solution} 

To integrate Eqs.~(\ref{eqAB}) we decompose $A$ and $B$ in Legendre polynomials, 
\begin{equation} 
A(r,\theta) = \sum_{\ell=0}^\infty A_\ell(r) P_\ell(\cos\theta), \qquad 
B(r,\theta) = \sum_{\ell=0}^\infty B_\ell(r) P_\ell(\cos\theta), 
\label{AB_series} 
\end{equation} 
and we write 
\begin{equation} 
\delta(\cos\theta - 1) = \frac{1}{2} \sum_{\ell=0}^\infty (2\ell+1) P_\ell(\cos\theta);  
\end{equation} 
this identity follows from the completeness relation for spherical harmonics, after an average over $\phi$. The equations become 
\begin{subequations}
\label{eqABL}
\begin{align} 
r^2 f \frac{d^2 A_\ell}{dr^2} + 2(r-M) \frac{dA_\ell}{dr} - \biggl[ \ell(\ell+1) + \frac{L^2}{r^2 f} \biggr] A_\ell 
&= -(2\ell+1) \Gamma_A\, \delta(r-r_0)\, 
\label{eqAL} \\ 
r^2 f \frac{d^2 B_\ell}{dr^2} + 2(r-M) \frac{dB_\ell}{dr} - \ell(\ell+1) B_\ell 
&= -(2\ell+1) \Gamma_B\, \delta(r-r_0). 
\label{eqBL} 
\end{align} 
\end{subequations} 
A change of variable from $r$ to $\xi := (r-M)/L$ brings the equations to the form 
\begin{subequations}
\label{eqABLxi}
\begin{align} 
(\xi^2-1) A_\ell'' + 2\xi A_\ell' - \biggl[ \ell(\ell+1) + \frac{1}{\xi^2-1} \biggr] A_\ell 
&= -(2\ell+1) \frac{\Gamma_A}{L}\, \delta(\xi - \xi_0),  
\label{eqALxi} \\ 
(\xi^2-1) B_\ell'' + 2\xi B_\ell' - \ell(\ell+1) B_\ell 
&= -(2\ell+1) \frac{\Gamma_B}{L}\, \delta(\xi - \xi_0),  
\label{eqBLxi} 
\end{align} 
\end{subequations} 
in which a prime indicates differentiation with respect to $\xi$, and $\xi_0 := (r_0-M)/L$. When $\xi \neq \xi_0$ the solution to Eq.~(\ref{eqALxi}) is a combination of $P_\ell^1 = (\xi^2-1)^{1/2} P_\ell'$ and $Q_\ell^1 = (\xi^2-1)^{1/2} Q_\ell'$; the solution to Eq.~(\ref{eqBLxi}) is a combination of $P_\ell$ and $Q_\ell$. To account for the delta function on the right-hand side of the equations, we must ensure that $A_\ell$ and $B_\ell$ are continuous at $\xi = \xi_0$, but that $A_\ell'$ and $B_\ell'$ possess the correct discontinuity. 

It is useful to note that the event horizon is situated at $\xi = 1$, and that the solutions to Eqs.~(\ref{eqABLxi}) are defined in the interval $1 \leq \xi < \infty$. The functions $f^{1/2} A_\ell$ and $B_\ell$ must be regular at $\xi = 1$ and go to zero at $\xi = \infty$; we have that $r^2 f = L^2(\xi^2-1)$, and so we demand that $(\xi^2-1)^{1/2} A_\ell$ be regular at $\xi = 1$. For $A_\ell$, the physically relevant solution to Eq.~(\ref{eqALxi}) is proportional to $P_\ell^1$ when $\xi < \xi_0$, and to $Q_\ell^1$ when $\xi > \xi_0$. For $B_\ell$, the relevant solution to Eq.~(\ref{eqBLxi}) is proportional to $P_\ell$ when $\xi < \xi_0$, and to $Q_\ell$ when $\xi > \xi_0$. 

Taking into account the junction conditions at $\xi = \xi_0$, we find that the global solutions to Eqs.~(\ref{eqABLxi}) are 
\begin{equation} 
A_\ell = -\frac{2\ell+1}{\ell(\ell+1)} \frac{\Gamma_A}{L} (\xi_0^2-1)^{1/2} (\xi^2-1)^{1/2} \left\{
\begin{array}{ll} 
Q_\ell'(\xi_0) P_\ell'(\xi) & \quad \xi < \xi_0 \\ 
P_\ell'(\xi_0) Q_\ell'(\xi) & \quad \xi > \xi_0 
\end{array} \right.  
\label{ALsol} 
\end{equation} 
and  
\begin{equation} 
B_\ell = (2\ell+1) \frac{\Gamma_B}{L} \left\{
\begin{array}{ll} 
Q_\ell(\xi_0) P_\ell(\xi) & \quad \xi < \xi_0 \\ 
P_\ell(\xi_0) Q_\ell(\xi) & \quad \xi > \xi_0 
\end{array} \right. . 
\label{BLsol} 
\end{equation} 
Equation (\ref{ALsol}) does not apply to $\ell = 0$, and this case requires a separate treatment. When $\ell = 0$ the two independent solutions to Eq.~(\ref{eqALxi}) are $(\xi^2-1)^{-1/2}$ and $\xi(\xi^2-1)^{-1/2}$. The first solution vanishes at infinity, and both solutions are regular at the horizon. The global solution is chosen to be 
\begin{equation} 
A_0 = \frac{\Gamma_A}{L} (\xi_0^2-1)^{-1/2} (\xi^2-1)^{-1/2} \left\{
\begin{array}{ll} 
\xi & \quad \xi < \xi_0 \\ 
\xi_0 & \quad \xi > \xi_0 
\end{array} \right. . 
\label{A0sol} 
\end{equation}
This will be seen below to require an adjustment.  

\subsection{Potentials in closed form} 
\label{subsec:potentials} 

The summation identities --- Eqs.~(110) and (34) of Ref.~\cite{bini-geralico-ruffini:07}, respectively ---
\begin{equation} 
\frac{1}{\sqrt{x^2 - 2xy\cos\theta + y^2 - \sin^2\theta}} 
= \sum_{\ell=0}^\infty (2\ell+1) 
\left\{ \begin{array}{c}
Q_\ell(y) P_\ell(x) \\ 
P_\ell(y) Q_\ell(x) 
\end{array} \right\} P_\ell(\cos\theta) 
\label{SF1} 
\end{equation} 
and 
\begin{equation} 
\frac{xy-\cos\theta}{\sqrt{x^2 - 2xy\cos\theta + y^2 - \sin^2\theta}} 
= 
\left\{ \begin{array}{c}
x \\
y 
\end{array} \right\} 
- (x^2-1)(y^2-1) \sum_{\ell=1}^\infty \frac{2\ell+1}{\ell(\ell+1)}  
\left\{ \begin{array}{c}
Q'_\ell(y) P'_\ell(x) \\ 
P'_\ell(y) Q'_\ell(x) 
\end{array}
\right\} P_\ell(\cos\theta), 
\label{SF2} 
\end{equation} 
are derived in Appendix E of Ref.~\cite{lahaye-poisson:21}; the upper row refers to the case $x < y$, while the lower row refers to $x > y$. They allow us to sum the series of Eqs.~(\ref{AB_series}) and express $A$ and $B$ in closed forms.  Setting $x = \xi$ and $y = \xi_0$, we find that Eqs.~(\ref{ALsol}), (\ref{A0sol}), and (\ref{SF2}) give
\begin{equation} 
A_{\rm p} = \frac{\Gamma_A}{L (\xi_0^2-1)^{1/2}} 
\frac{\xi_0 \xi - \cos\theta}{(\xi^2-1)^{1/2}
(\xi^2-2\xi_0\xi\cos\theta + \xi_0^2 - \sin^2\theta)^{1/2}}. 
\end{equation} 
On the other hand, Eqs.~(\ref{BLsol}) and (\ref{SF1}) return 
\begin{equation} 
B_{\rm p} = \frac{\Gamma_B}{L} \frac{1}{(\xi^2-2\xi_0\xi\cos\theta + \xi_0^2 - \sin^2\theta)^{1/2}}. 
\end{equation} 
We place a label ``p'' on the functions to indicate that these are particular solutions to the field equations for a point charge at $r=r_0$. As we shall see, these solutions must be amended to ensure that all boundary conditions are satisfied. Inserting $\xi = (r-M)/L$ and $\xi_0 = (r_0-M)/L$ into these expressions, we arrive at 
\begin{subequations}
\label{ABfinal} 
\begin{align}  
\sqrt{f} A_{\rm p} &= \frac{k_A}{r_0} \frac{(r_0-M)(r-M) - L^2\cos\theta}{rD}, \\
B_{\rm p} &= \frac{k_B}{D}, 
\end{align}
\end{subequations} 
where 
\begin{subequations} 
\label{kdef} 
\begin{align} 
k_A &:= \frac{1}{L^2} \Bigl[ q(M-Q^2/r_0) - m Q \sqrt{f_0} \Bigr], \\ 
k_B &:= \frac{\sqrt{f_0}}{L^2} \Bigl[ m(M-Q^2/r_0) - q Q \sqrt{f_0} \Bigr]
\end{align} 
\end{subequations} 
and 
\begin{equation} 
D := \bigl[(r-M)^2 - 2(r_0-M)(r-M)\cos\theta + (r_0-M)^2 - L^2\sin^2\theta \bigr]^{1/2} 
\label{D_def} 
\end{equation} 
is the spatial distance in the RN spacetime between a point at $(r,\theta)$ and the particle at $(r_0, 0)$. We note that $k_A = \Gamma_A/\sqrt{f_0}$ and $k_B = \Gamma_B$, and recall that $L^2 := M^2-Q^2$. The original potentials are then recovered from Eq.~(\ref{decoupling}),  
\begin{equation} 
\Phi_{\rm p} = (M - Q^2/r) (\sqrt{f} A_{\rm p} )+ Q f B_{\rm p}, \qquad 
U_{\rm p} = (M-Q^2/r) B_{\rm p} + Q (\sqrt{f} A_{\rm p}), 
\label{particular} 
\end{equation} 
where we again indicate that these are particular solutions.

\subsection{Gauss' law} 

The potentials $\Phi_{\rm p}$ and $U_{\rm p}$ describe a situation in which a particle of mass $m$ and charge $q$ has been added to the spacetime, but in which the black hole has also acquired an additional charge. To see this, we appeal to Gauss' law, which states that the total charge enclosed by a two-surface $S$ is given by 
\begin{equation} 
{\sf Q}(S) = -\frac{1}{4\pi} \oint_S F_{\alpha\beta} n^\alpha r^\beta\, dS, 
\label{gauss_law} 
\end{equation} 
where $n^\alpha$ is the surface's timelike unit normal (pointing to the future), $r^\alpha$ is its spacelike unit normal (pointing out of $S$), and $dS$ is the induced surface element. We choose $S$ to be a surface of constant $t$ and $r$, set
\begin{equation} 
n^\alpha = e^{U} f^{-1/2}[1, 0, 0, 0], \qquad 
r^\alpha = e^{-(U+\gamma)} f^{1/2} [0, 1, 0, 0], \qquad 
dS = e^{(2U+\gamma)} r^2\sin\theta\, d\theta d\phi,
\end{equation} 
and obtain  
\begin{equation} 
{\sf Q}(S) = Q -\frac{1}{2} \int_0^\pi \bigl( r^2 \partial_r \Phi - 2 Q U \bigr)\sin\theta\, d\theta. 
\end{equation} 
To evaluate this we follow Ref.~\cite{bini-geralico-ruffini:07} and write $f^{1/2} A_{\rm p} = (k_A/r_0 r) (\partial_\theta D/\sin\theta)$. Making the substitution in $\Phi_{\rm p}$ and $U_{\rm p}$, we find that 
\begin{equation} 
-\frac{1}{2} \bigl( r^2 \partial_r \Phi_{\rm p} - 2 Q U_{\rm p} \bigr)\sin\theta 
= \frac{k_A}{2 r_0} \Bigl[ -(Mr-Q^2) \partial_{r\theta} D 
+ M \partial_\theta D \Bigr] - \frac{1}{2} k_B Q r^2 f (\partial_r D^{-1})\, \sin\theta. 
\end{equation} 
Integration is accomplished with
\begin{subequations}
\begin{align} 
D(\pi) - D(0) &= r + r_0 - 2M - (r-r_0) \sgn(r - r_0), \\ 
\partial_r [ D(\pi) - D(0) ] &= 1 - \sgn(r - r_0), \\
r^2 f \int_0^\pi (\partial_r D^{-1})\sin\theta\, d\theta &= -1 - \sgn(r-r_0),
\end{align}
\end{subequations} 
where $\sgn(r-r_0) $ is equal to $+1$ when $r > r_0$ and $-1$ when $r < r_0$. We obtain 
\begin{equation} 
{\sf Q}(r < r_0) = Q - \frac{k_A L^2}{r_0}, \qquad 
{\sf Q}(r > r_0) = Q + k_A M \biggl(1 - \frac{M}{r_0} \biggr) + k_B Q. 
\end{equation} 
We see that indeed, the black hole has acquired a charge $-k_A L^2/r_0$, and that the charge at infinity is not equal to $Q + q$, as it should. 

\subsection{Monopole potentials} 

To restore the correct physical situation, we must add to $\Phi_{\rm p}$ and $U_{\rm p}$ potentials that shift the charge of the black hole back to $Q$. These monopole solutions to the homogeneous field equations are constructed in Appendix~\ref{sec:monopole}, and they are given by 
\begin{equation} 
\Phi_{\rm mono} = \frac{k_A}{r_0} \biggl( \frac{M^2+Q^2}{r} - \frac{MQ^2}{r^2} \biggr), \qquad
U_{\rm mono} = -\frac{k_A Q}{r_0} \biggl( 1 - \frac{M}{r} \biggr). 
\label{mono} 
\end{equation} 
With these amendments, the correct solutions to the field equations are 
\begin{equation} 
\Phi = \Phi_{\rm p} + \Phi_{\rm mono}, \qquad 
U = U_{\rm p} + U_{\rm mono}. 
\label{complete_slns} 
\end{equation} 
Recalculating ${\sf Q}(S)$ with the amended potentials, we now find that 
\begin{equation} 
{\sf Q}(r < r_0) = Q, \qquad  
{\sf Q}(r > r_0) = Q + q;  
\end{equation} 
we made use of Eqs.~(\ref{kdef}) to simplify the expression for ${\sf Q}(r>r_0)$. The black hole now has a charge $Q$, and the charge at infinity is $Q + q$. All is good. 

A remarkable aspect of our solution is that $U$ does not go to zero when $r \to \infty$. Instead we have that
\begin{equation}
U_\infty := U(r \to \infty) = U_{\rm mono}(r \to \infty) 
= -\frac{k_A Q}{r_0}. 
\label{U_infinity}
\end{equation}
This implies that in the perturbed spacetime, the time coordinate $t$ no longer measures proper time for a static observer at infinity. This role is taken over by the rescaled time $\hat{t} = e^{-U_\infty} t$. This observation will have consequences in subsequent calculations; care will be required to account for the nontrivial asymptotic behavior of the metric.

\subsection{Komar mass} 

To further check the validity of our solution, we calculate the mass at infinity, as defined by the Komar integral 
\begin{equation} 
{\sf M}(S) = \frac{1}{4\pi} \oint_S (\nabla_\alpha \hat{t}_\beta) n^\alpha r^\beta\, dS, 
\label{komar_mass} 
\end{equation} 
in which $\hat{t}^\alpha = e^{U_\infty}[1,0,0,0]$ is the spacetime's timelike Killing vector, properly normalized at infinity. We are interested in the limit in which $S$ is pushed out to $r = \infty$. Performing the computation, we obtain 
\begin{equation} 
{\sf M}(\infty) = M + k_A Q \biggl(1 - \frac{M}{r_0} \biggr) + k_B M, 
\end{equation} 
or 
\begin{equation} 
{\sf M}(\infty) = M + m \sqrt{f_0} + \frac{qQ}{r_0}   
\end{equation} 
after inserting Eqs.~(\ref{kdef}). The difference ${\sf M}(\infty) - M$ is recognized as the particle's Killing energy in the background RN spacetime, as given by Eq.~(\ref{killing_energy}). Because the total mass is correctly recovered, we conclude that our amended potentials give a faithful description of the spacetime.  

\subsection{Calculation of $\gamma$}
\label{subsec:g1_calc} 

The calculation of $\gamma$ proceeds on the basis of Eqs.~(\ref{g_eqns}), in which we insert $U$ and $\Phi$ from Eq.~(\ref{complete_slns}). The general solution for $\gamma$ features a constant of integration, and we choose it so that $\gamma(r,\theta=0) = 0$ when $M+L < r < r_0$, that is, between the black hole and particle. With this choice we find that
\begin{equation} 
\gamma = \frac{2k_B L^2}{r_0^2 f_0} \biggl[ \frac{r-M - (r_0-M)\cos\theta}{D} + 1 \biggr],
\label{gamma1}
\end{equation}
where $k_B$ is defined by Eq.~(\ref{kdef}) and $D$ by Eq.~(\ref{D_def}). With this solution we have that $\gamma(r,\theta=0) = 4F$ when $r > r_0$ (above the particle), with $F$ denoting the force of Eq.~(\ref{force}). We also have that $\gamma(r,\theta=\pi) = 4F$ for any value of $r$.  

The nonzero value of $\gamma$ on the axis, above the particle on the upper axis (at $\theta = 0$), and everywhere on the lower axis (at $\theta = \pi$), implies that the spacetime comes with an angular deficit: the ratio of proper circumference to proper radius for a small circle around the axis is not equal to $2\pi$, as required by elementary flatness. The conical singularity betrays the presence of a thin distribution of matter on the axis --- a massless string --- with a tension $T$ equal to its linear mass density $\mu$. The string on the upper axis is the external agent that keeps the particle in place at $r=r_0$ and prevents it from falling toward the black hole; the string tension $T$ is equal to the force $F$ acting on the particle. Similarly, the string on the lower axis is attached to the black hole and prevents it from falling toward the particle; the tension is the same in both strings. (For an extended discussion of these properties, please refer to Ref.~\cite{lahaye-poisson:21}.)

With a different choice of constant of integration we could arrange for $\gamma(r,\theta=0)$ to be nonzero between the black hole and particle, and zero below the black hole and above the particle. In this case the conical singularity would lie between the black hole and particle, and it would signal the presence of a strut between the objects; the strut would then be responsible for keeping them apart. With yet another choice of constant there would be a conical singularity everywhere on the axis. In this case the spacetime would contain a strut and two strings, and all these would share the responsibility of keeping the objects apart. The choice of constant becomes immaterial when the situation is specialized to the one of
Sec.~\ref{sec:pert-equil}. When the condition of Eq.~(\ref{equil-condition}) is imposed we shall find that the conical singularity disappears altogether, whether it was below the black hole, above the particle, or between the objects. The choice of constant, therefore, has no incidence on the vanishing of tidally induced multipole moments for black holes. 

\subsection{Properties of the black hole}

We use the metric of Eq.~(\ref{metric}) to calculate how the black hole is affected by the perturbation. The perturbed event horizon continues to be situated at $r = M+L$, where $g_{tt} = 0$, and the induced metric on the horizon is given by the $\theta$-$\theta$ and $\phi$-$\phi$ components of the metric, with $U$ and $\gamma$ evaluated at $r=M+L$. The horizon values are
\begin{equation}
U(r=M+L) = \frac{k_B L}{r_0-M-L\cos\theta}, \qquad
\gamma(r=M+L) = \frac{2k_B L^2}{r_0-M-L} \frac{1-\cos\theta}{r_0-M-L\cos\theta},
\end{equation}
and we also have that $\Phi(r=M+L) = k_AM/r_0$; to arrive at our expression for $\gamma$ we made use of the factorization $r_0^2 f_0 = (r_0-M-L)(r_0-M+L)$. With this information we find that the black-hole area is 
\begin{equation}
{\cal A} = 4\pi (M+L)^2 \biggl( 1 + \frac{2k_B L}{r_0-M-L} \biggr). 
\label{horizon_area} 
\end{equation} 

The surface gravity follows from $\kappa^2 = -\frac{1}{2} (\nabla_\alpha \hat{t}_\beta) (\nabla^\alpha \hat{t}^\beta)$, where $\hat{t}^\alpha$ is the rescaled timelike Killing vector, and where the right-hand side is evaluated on the horizon. We find that
\begin{equation}
\kappa = \frac{L}{(M+L)^2} \biggl( 1 - \frac{k_AQ}{r_0} - \frac{2k_B L}{r_0-M-L} \biggr); 
\label{surface_gravity}
\end{equation}
in accordance with the zeroth law of black-hole mechanics, the surface gravity is constant on the horizon.

The horizon-valued electrostatic potential $\Psi_{\rm H}$ must also be defined with respect to the rescaled Killing vector. We have $\Psi_{\rm H} := -A_\alpha \hat{t}^\alpha = e^{U_\infty} ( Q/r + \Phi \bigr)$, where the right-hand side is evaluated at $r=M+L$. A short calculation returns
\begin{equation}
\Psi_{\rm H} = \frac{Q}{M+L} + \frac{k_A L}{r_0}.
\end{equation}

The Smarr mass of the black hole is defined by
\begin{equation}
M_{\rm Smarr} := \frac{\kappa {\cal A}}{4\pi} + Q \Psi_{\rm H},
\end{equation} 
and we find that all perturbative terms cancel out, leaving 
\begin{equation}
M_{\rm Smarr} = M.
\label{Smarr_mass} 
\end{equation}
The mass parameter $M$ of the perturbed black hole can therefore be related to geometric objects defined on the event horizon.  

\section{Perturbed spacetime: Equilibrium case} 
\label{sec:pert-equil}

We specialize the results of the preceding section to a particle with a mass $m$ and charge $q$ related by Eq.~(\ref{ratio}), copied here for ease of reference:
\begin{equation}
m(M - Q^2/r_0) = q Q \sqrt{f_0}. 
\label{equil-condition}
\end{equation}
The particle's charge-to-mass ratio is therefore restricted to be a specific function of $r_0$. When this condition is satisfied we find from Eq.~(\ref{force}) that $F = 0$, where $F$ is the force required of an external agent to hold the particle in place. The particle, therefore, is a state of equilibrium that results from equal and opposite gravitational and electrostatic forces. 

Another consequence of Eq.~(\ref{equil-condition}) is that the parameters defined by Eq.~(\ref{kdef}) are now given by 
\begin{equation}
k_A = \frac{m}{Q\sqrt{f_0}}, \qquad k_B = 0. 
\end{equation}
The vanishing of $k_B$ implies the remarkable fact that $\gamma = 0$, where $\gamma$ is the potential of Eq.~(\ref{gamma1}). The perturbed spacetime no longer features a conical singularity, and strings are no longer required to hold the particle and black hole.

The potentials reduce to 
\begin{subequations}
\label{UP-equil}
\begin{align}
\Phi &= \frac{m}{Q\, r_0 \sqrt{f_0}} \Biggl[ \biggl( M- \frac{Q^2}{r} \biggr)
\frac{(r_0-M)(r-M) - L^2 \cos\theta}{rD}   
+ \frac{M^2+Q^2}{r} - \frac{M Q^2}{r^2} \Biggr], \\
U &= \frac{m}{r_0 \sqrt{f_0}} \biggl[ \frac{(r_0-M)(r-M) - L^2\cos\theta}{r D}
- 1 + \frac{M}{r} \biggr]. 
\end{align}
\end{subequations}
These include the particular solutions of Eq.~(\ref{particular}) and the monopole contributions of Eq.~(\ref{mono}). We recall that $L^2 := M^2-Q^2$ and that $D$ is given by Eq.~(\ref{D_def}).
The potentials become
\begin{equation}
\Phi_{\rm H} = \frac{mM}{Q r_0 \sqrt{f_0}}, \qquad
U_{\rm H} = 0
\label{pot_horizon} 
\end{equation}
when evaluated on the horizon. 

Because $\gamma = 0$ in this equilibrium situation, the perturbed metric simplifies to 
\begin{equation} 
ds^2 = -e^{-2U} f\, dt^2 + e^{2U} \bigl[ f^{-1}\, dr^2
+ r^2 (d\theta^2 + \sin^2\theta\, d\phi^2) \bigr]. 
\label{metric-equil} 
\end{equation}
The vanishing of $U$ at $r = M + L$ implies that the intrinsic geometry of the event horizon is spherically symmetric, in spite of the tidal deformation of the surrounding spacetime. There is no direct link between the spherical state of the horizon and the vanishing of the tidally induced multipole moments of the black hole; these, we recall, are a property of the gravitational field at a large distance from the body.

{\it The coordinates of Eq.~(\ref{harm_coordinates}) continue to be harmonic in the perturbed spacetime.} This remarkable observation is verified by showing that each one of the scalar fields $X^\mu$ satisfies the wave equation of Eq.~(\ref{wave-eqn}), not just in the background spacetime defined by the RN metric, but also in the perturbed spacetime defined by the metric of Eq.~(\ref{metric-equil}). The statement is true irrespective of the precise form of the perturbation $U(r,\theta)$; the potential does not even appear in the perturbed wave equation. The statement, however, relies sensitively on the vanishing of $\gamma$; it would not be true otherwise.

It is a simple matter to use Eq.~(\ref{harm_coordinates}) to transform the perturbed metric to the harmonic coordinates $(X^0 = t, X^a)$, and from this to compute the gothic inverse metric $\gothg^{\alpha\beta} := \sqrt{-g}\, g^{\alpha\beta}$. We find that its time-time component is given by 
\begin{equation}
W := -\gothg^{tt} = e^{4U} \frac{(1 + M/\bar{r})^4}{1 - L^2/\bar{r}^2},
\label{gothg-equil}
\end{equation} 
where $\bar{r} = r - M$ is the harmonic radial coordinate. The multipole decomposition of $W$ is obtained by combining Eqs.~(\ref{decoupling}), (\ref{AB_series}), (\ref{ALsol}), (\ref{A0sol}), and (\ref{mono}). We reintroduce $\xi := \bar{r}/L$, $\xi_0 := \bar{r}_0/L$, set $\mu := M/L$, and write
\begin{equation}
W = W_{\rm RN} + \sum_{\ell=0}^\infty W_\ell(\xi)\, P_\ell(\cos\theta), 
\label{W_decomposed} 
\end{equation}
where 
\begin{equation}
W_{\rm RN} = \frac{(\xi+\mu)^4}{\xi^2(\xi^2-1)}
\end{equation}
is the Reissner-Nordstr\"om piece,  
\begin{subequations}
\label{Win} 
\begin{align}
W_0 &= 0, \\
W_\ell &= -\frac{4(2\ell+1)}{\ell(\ell+1)} \frac{m}{L} (\xi_0^2-1)^{1/2} Q_\ell'(\xi_0)\,
\frac{(\xi+\mu)^3}{\xi^2} P_\ell'(\xi)
\end{align}
\end{subequations}
when $\xi \leq \xi_0$, while 
\begin{subequations}
\label{Wout}
\begin{align}
W_0 &= -4 \frac{m}{L} (\xi_0^2-1)^{-1/2}\,
\frac{(\xi+\mu)^3 (\xi-\xi_0)}{\xi^2 (\xi^2-1)}, \\
W_\ell &= -\frac{4(2\ell+1)}{\ell(\ell+1)} \frac{m}{L} (\xi_0^2-1)^{1/2} P_\ell'(\xi_0)\,
\frac{(\xi+\mu)^3}{\xi^2} Q_\ell'(\xi) 
\end{align}
\end{subequations} 
when $\xi \geq \xi_0$. Equation (\ref{Win}) was already previewed in Eq.~(\ref{Well}), and described in some detail back in Sec.~\ref{sec:BHstatic}. The dipole contribution to Eq.~(\ref{W_decomposed}) is discussed in Appendix~\ref{sec:dipole}. 

\begin{acknowledgments} 
I am grateful to Michael LaHaye for valuable discussions. This work was supported by the Natural Sciences and Engineering Research Council of Canada.  
\end{acknowledgments} 

\appendix

\section{Monopole potentials} 
\label{sec:monopole} 

In this appendix we construct monopole solutions to the homogeneous version of Eqs.~(\ref{UP_eqns}). We insert $A=A(r)$ and $B=B(r)$ within Eqs.~(\ref{eqA}) and (\ref{eqB}), set the right-hand sides to zero, integrate the equations, and substitute the solutions within Eq.~(\ref{decoupling}). The most general solution comes with four arbitrary constants. One of these gives rise to an irrelevant constant term in $\Phi$, and another constant multiplies a function that diverges on the event horizon. The physically relevant solution therefore contains two arbitrary constants, denoted $a_1$ and $a_2$, and it is given by 
\begin{equation} 
\Phi_{\rm mono} = a_1 \biggl( \frac{M}{r} - \frac{Q^2}{r^2} \biggr) + a_2 \frac{Q}{r}, \qquad 
U_{\rm mono} = a_1 \frac{Q}{r} - a_2. 
\end{equation} 
Equations (\ref{g_eqns}) imply that $\gamma$ vanishes for this solution. 

It is intuitively clear that this two-parameter family of solutions represents a shift in the mass and charge parameters of the Reissner-Nordstr\"om (RN) black hole. However, the fact that $U_{\rm mono}$ tends toward $-a_2$ (instead of zero) as $r \to \infty$ implies that the interpretation of $a_1$ and $a_2$ is subtle. It is wrong, in particular, to interpret $a_1 M + a_2 Q$ as a shift in charge, and $a_1 Q$ as a shift in mass.  

To arrive at a proper interpretation for $a_1$ and $a_2$, we transform the metric of Eq.~(\ref{metric}), and the vector potential of Eq.~(\ref{vector_potential}), to the standard RN form. In these manipulations we set $U = U_{\rm mono}$, $\Phi = \Phi_{\rm mono}$, $\gamma = 0$, and work to first order in both $a_1$ and $a_2$. 

The first step is to introduce a new radial coordinate $\hat{r} = e^{U} r = (1 + U) r$, which converts the angular part of the line element to the standard $\hat{r}^2 (d\theta^2 + \sin^2\theta\, d\phi^2)$. Inserting $U$ and inverting, we have that    
\begin{equation}  
r = (1+a_2) \hat{r} - a_1 Q. 
\end{equation} 
The second step is to express $g_{tt}$ and $g_{\hat{r}\hat{r}}$ in terms of the new radial coordinate. Simple algebra returns $g_{tt} = -(1+2a_2) \hat{f}$ and $g_{\hat{r}\hat{r}} = 1/\hat{f}$, where 
\begin{equation} 
\hat{f} = 1 - \frac{2(M+\delta M)}{\hat{r}} + \frac{(Q+\delta Q)^2}{\hat{r}^2}, 
\end{equation} 
with 
\begin{equation} 
\delta M := a_1 Q - a_2 M, \qquad \delta Q := a_1 M - a_2 Q. 
\label{shifts} 
\end{equation} 
The third step is to eliminate the factor of $1 + 2a_2$ in $g_{tt}$, by introducing a rescaled time coordinate $\hat{t}$ such that 
\begin{equation} 
t = (1-a_2) \hat{t}. 
\end{equation} 
With all this the metric becomes 
\begin{equation} 
ds^2 = -\hat{f}\, d\hat{t}^2 + \hat{f}^{-1}\, d\hat{r}^2 + \hat{r}^2 (d\theta^2 + \sin^2\theta\, d\phi^2), 
\end{equation} 
which is the standard form of the RN metric. The fourth and final step is to re-express the vector potential in terms of $\hat{t}$ and $\hat{r}$. We obtain 
\begin{equation} 
A_\alpha = -\frac{Q+\delta Q}{\hat{r}}\, \partial_\alpha \hat{t}. 
\end{equation} 
The transformation reveals that we are indeed dealing with another RN solution, with a new mass $M + \delta M$ and a new charge $Q + \delta Q$. The shifts in the mass and charge parameters are given by Eq.~(\ref{shifts}). 

This interpretation is confirmed with an application of Gauss' law --- Eq.~(\ref{gauss_law}) --- and a computation of the Komar mass of Eq.~(\ref{komar_mass}). Working in the original $(t,r,\theta,\phi)$ coordinates, and choosing again $S$ to be a two-surface of constant $t$ and $r$, we find that 
\begin{equation} 
{\sf Q}(S) = Q + a_1 M - a_2 Q 
\end{equation} 
and 
\begin{equation} 
{\sf M}(S) = (1-a_2) \biggl( M - \frac{Q^2}{r} \biggr) + a_1 Q f. 
\end{equation} 
The first equation reveals once more that $\delta Q = a_1 M - a_2 Q$, and the second equation shows that the mass at infinity is $M + \delta M$ with $\delta M = a_1 Q - a_2 M$. 

Equation (\ref{shifts}) reveals that the choice
\begin{equation} 
a_1 = -\frac{Q}{M^2-Q^2}\, \delta M, \qquad 
a_2 = -\frac{M}{M^2-Q^2}\, \delta M 
\end{equation} 
produces potentials that correspond to a pure mass shift, accompanied by $\delta Q = 0$. These are 
\begin{equation} 
\Phi_{\rm mass} = -\frac{Q\, \delta M}{M^2-Q^2} \biggl( \frac{2M}{r} - \frac{Q^2}{r^2} \biggr), \qquad 
U_{\rm mass} = \frac{\delta M}{M^2 - Q^2} \biggl( M - \frac{Q^2}{r} \biggr). 
\end{equation} 
On the other hand, the assignment
\begin{equation} 
a_1 = \frac{M}{M^2-Q^2}\, \delta Q, \qquad 
a_2 = \frac{Q}{M^2-Q^2}\, \delta Q 
\end{equation} 
gives rise to a pure charge shift, accompanied by $\delta M = 0$. The corresponding potentials are 
\begin{equation} 
\Phi_{\rm charge} = \frac{\delta Q}{M^2 - Q^2} \biggl( \frac{M^2+Q^2}{r} - \frac{MQ^2}{r^2} \biggr), \qquad 
U_{\rm charge} = -\frac{Q\, \delta Q}{M^2-Q^2} \biggl( 1 - \frac{M}{r} \biggr). 
\end{equation} 

The monopole potentials introduced in Eq.~(\ref{mono}) come with the shifts $\delta M = 0$ and $\delta Q = k_A (M^2-Q^2)/r_0$.  

\section{Dipole contribution to $W$} 
\label{sec:dipole} 

In this appendix we examine the dipole contribution to $W := -\gothg^{tt}$ and explain its significance. From Eqs.~(\ref{Win}) and (\ref{Wout}) we have that
\begin{equation}
W_1 = \frac{3m}{L} (\xi_0^2-1)^{1/2} \biggl[ \ln\frac{\xi_0-1}{\xi_0+1}
+ \frac{2\xi_0}{\xi_0^2-1} \biggr] \frac{(\xi+\mu)^3}{\xi^2}
\label{W1_in} 
\end{equation}
when $\xi < \xi_0$, and
\begin{equation}
W_1 = \frac{3m}{L} (\xi_0^2-1)^{1/2} 
\biggl[ \ln\frac{\xi-1}{\xi+1} + \frac{2\xi}{\xi^2-1} \biggr] \frac{(\xi+\mu)^3}{\xi^2}
\label{W1_out} 
\end{equation}
when $\xi > \xi_0$.

The factor involving $\xi$ in Eq.~(\ref{W1_in}) can be expanded, and the term with the smallest power of $\xi$ is $\mu^3/\xi^2$. In light of the considerations of Secs.~\ref{sec:compact} and \ref{sec:BHstatic}, this indicates that when viewed as a skeletonized post-Newtonian object, the black hole possesses a mass dipole moment. This, in turn, implies that the spatial origin of the harmonic coordinates is not anchored at the black hole's center of mass. 

Similarly, the spatial origin of the harmonic coordinates is not attached to the center of mass of the entire system, black hole plus particle. This can be seen from Eq.~(\ref{W1_out}) and the fact that the factor involving $\xi$ becomes approximately $(4/3) \xi^{-2}$ when $\xi \gg 1$. This shows that the system possesses a nonvanishing dipole moment at infinity.

It is by no means necessary for the spatial origin of the coordinate system to coincide with the system's center of mass, and the properties reviewed above do not constitute a problem. It is instructive, nevertheless, to introduce a gauge transformation that produces such a coincidence. Before we proceed we recall from Eqs.~(\ref{decoupling}), (\ref{AB_series}) and (\ref{ALsol}) that the dipole piece of the potential $U(r,\theta)$ is given by $U_1(r) \cos\theta$, where
\begin{equation}
U_1 = \frac{3m}{4L^3} \biggl[ r_0 \sqrt{f_0} \ln\frac{r_0-M-L}{r_0-M+L}
+ \frac{2L(r_0-M)}{r_0\sqrt{f_0}} \biggr] r f
\label{U1_in} 
\end{equation}
when $r < r_0$, and
\begin{equation}
U_1 = \frac{3m}{4L^3} r_0 \sqrt{f_0} \biggl[ r f \ln\frac{r-M-L}{r-M+L}
+ 2 L \biggl(1 - \frac{M}{r} \biggr) \bigg] 
\label{U1_out} 
\end{equation}
when $r > r_0$. For the purpose of this discussion we reinstate the original Reissner-Nordstr\"om coordinates. 

We wish to find a gauge transformation that makes $U_1(r)$ vanish when $r > r_0$; this will remove the dipole moment at infinity and anchor the new coordinates at the center of mass of the entire system. The procedure can easily be adapted to make $U_1(r)$ vanish when $r < r_0$ instead; this would attach the new coordinates to the black hole's center of mass.

The transformation is generated by the gauge vector
\begin{equation}
\Xi_\alpha = \bigl[ 0, h(r) \cos\theta, -j(r) \sin\theta, 0 \bigr],
\end{equation}
where $h(r)$ and $j(r)$ are functions to be determined. The transformation produces a change $\Delta p_{\alpha\beta} = -\nabla_\alpha \Xi_\beta - \nabla_\beta \Xi_\alpha$ in the metric perturbation $p_{\alpha\beta} := \delta g_{\alpha\beta}$, and the change is given explicitly by 
\begin{subequations}
\begin{align}
\Delta p_{tt} &= \frac{2}{r^2} f (M-Q^2/r) h \cos\theta, \\
\Delta p_{rr} &= -2 \biggl( \frac{dh}{dr} + \frac{M-Q^2/r}{r^2 f} h \biggr) \cos\theta, \\
\Delta p_{r\theta} &= \biggl( \frac{dj}{dr} - \frac{2}{r} j + h \biggr) \sin\theta, \\
\Delta p_{\theta\theta} &= -2 \bigl( r f h - j \bigr) \cos\theta,
\end{align}
\end{subequations}
and $\Delta p_{\phi\phi} = \Delta p_{\theta\theta}\, \sin^2\theta$. We set
\begin{equation}
h = -\frac{r^2}{M - Q^2/r} U_1, 
\end{equation}
with $U_1$ given by Eq.~(\ref{U1_out}), so that the new $p_{tt}$ vanishes when $r > r_0$. And we  choose $j$ to satisfy $dj/dr - 2 r^{-1} j + h = 0$, so that $p_{r\theta}$ continues to vanish; its explicit expression in terms of dilogarithms, logarithms, and rational functions is not terribly illuminating and is not displayed here.

The asymptotic behavior of the gauge functions when $r \to \infty$ is
\begin{equation}
h \sim -\frac{m}{M} r_0 \sqrt{f_0}, \qquad
j \sim -\frac{m}{M} r_0 \sqrt{f_0}\, r.
\end{equation}
Making the substitution in the gauge vector, and transforming to Cartesian coordinates $X^1 \sim r\sin\theta\cos\phi$, $X^2 \sim r\sin\theta\sin\phi$, and $X^3 \sim r\cos\theta$, we find that
\begin{equation}
\Xi^a \sim -\frac{m}{M} r_0 \sqrt{f_0}\, [0, 0, 1]. 
\end{equation}
As could be expected, the gauge transformation describes a translation along the symmetry axis.

The gauge transformation generated by $\Xi_\alpha$ does not preserve the harmonic property of the spatial coordinates $X^a$. The transformation, however, was devised to achieve an ambitious goal, to remove the dipole piece of $p_{tt}$ in the entire domain $r > r_0$. There is no need to be so ambitious. To eliminate the dipole moment at infinity it suffices to remove from $p_{tt}$ the leading term in an expansion in powers of $r^{-1}$. And this can be accomplished with the constant gauge vector
\begin{equation}
\Upsilon^a = -\frac{m}{M} r_0 \sqrt{f_0}\, [0, 0, 1]. 
\end{equation} 
Such a constant translation, $X^a \to X^a + \Upsilon^a$, does preserve the harmonic property of the spatial coordinates. 

\bibliography{/Users/poisson/writing/papers/tex/bib/master}

\begin{thebibliography}{10}
\expandafter\ifx\csname url\endcsname\relax
  \def\url#1{{\tt #1}}\fi
\expandafter\ifx\csname urlprefix\endcsname\relax\def\urlprefix{URL }\fi

\bibitem{flanagan-hinderer:08}
E.~E. Flanagan and T.~Hinderer, {\em Constraining neutron star tidal Love
  numbers with gravitational wave detectors\/}, Phys. Rev. D {\bf 77},
  021502(R) (2008), arXiv:0709.1915.

\bibitem{chatziioannou:20}
K.~Chatziioannou, {\em Neutron star tidal deformability and equation-of-state
  constraints\/}, Gen. Rel. Grav. {\bf 52}, 109 (2020), arXiv:2006.03168.

\bibitem{GW170817:17}
B.~P. Abbott {\em et~al.\/}, {\em GW170817: Observation of Gravitational Waves
  from a Binary Neutron Star Inspiral\/}, Phys. Rev. Lett. {\bf 119}, 161101
  (2017), arXiv:1710.05832.

\bibitem{GW170817:18}
B.~P. Abbott {\em et~al.\/}, {\em GW170817: Measurements of Neutron Star Radii
  and Equation of State\/}, Phys. Rev. Lett. {\bf 121}, 161101 (2018).

\bibitem{landry-essick-reed-chatziioannou:20}
P.~Landry, R.~Essick, and K.~Chatziioannou, {\em Nonparametric constraints on
  neutron star matter with existing and upcoming gravitational wave and pulsar
  observations\/}, Phys. Rev. D {\bf 101}, 123007 (2020), arXiv:2003.04880.

\bibitem{damour-nagar:09}
T.~Damour and A.~Nagar, {\em {Relativistic tidal properties of neutron
  stars}\/}, Phys. Rev. D {\bf 80}, 084035 (2009), arXiv:0906.0096.

\bibitem{binnington-poisson:09}
T.~Binnington and E.~Poisson, {\em Relativistic theory of tidal Love
  numbers\/}, Phys. Rev. D {\bf 80}, 084018 (2009), arXiv:0906.1366.

\bibitem{chirenti-posada-guedes:20}
C.~Chirenti, C.~Posada, and V.~Guedes, {\em Where is Love? Tidal deformability
  in the black hole compactness limit\/}, Class. Quantum Grav. {\bf 37}, 195017
  (2020).

\bibitem{hui-etal:21a}
L.~Hui, A.~Joyce, R.~Penco, L.~Santoni, and A.~R. Solomon, {\em Static response
  and Love numbers of Schwarzschild black holes\/}, JCAP {\bf 2021}(04), 052
  (2021).

\bibitem{chia:21}
H.~S. Chia, {\em Tidal deformation and dissipation of rotating black holes\/},
  Phys. Rev. D {\bf 104}, 024013 (2021).

\bibitem{charalambous-dubovsky-ivanov:21a}
P.~Charalambous, S.~Dubovsky, and M.~M. Ivanov, {\em On the vanishing of Love
  numbers for Kerr black holes\/}, J. High Energ. Phys. {\bf 2021}, 38 (2021).

\bibitem{porto:16}
R.~A. Porto, {\em The effective field theorist’s approach to gravitational
  dynamics\/}, Physics Reports {\bf 633}, 1--104 (2016), arXiv:1601.04914.

\bibitem{charalambous-dubovsky-ivanov:21b}
P.~Charalambous, S.~Dubovsky, and M.~M. Ivanov, {\em Hidden Symmetry of
  Vanishing Love Numbers\/}, Phys. Rev. Lett. {\bf 127}, 101101 (2021).

\bibitem{hui-etal:21b}
L.~Hui, A.~Joyce, R.~Penco, L.~Santoni, and A.~R. Solomon, {\em {Ladder
  symmetries of black holes: Implications for Love mumbers and no-hair
  theorems}\/}  (2021).

\bibitem{poisson-will:14}
E.~Poisson and C.~M. Will, {\em Gravity: Newtonian, Post-Newtonian,
  Relativistic\/} (Cambridge University Press, Cambridge, England, 2014).

\bibitem{gralla:18}
S.~E. Gralla, {\em {On the ambiguity in relativistic tidal deformability}\/},
  Class. Quant. Grav. {\bf 35}, 085002 (2018), arXiv:1710.11096.

\bibitem{geroch:70}
R.~Geroch, {\em Multipole moments. II. Curved space\/}, J. Math. Phys. {\bf
  11}, 2580--2588 (1970).

\bibitem{hansen:74}
R.~O. Hansen, {\em Multipole moments of stationary space‐times\/}, J. Math.
  Phys. {\bf 15}, 46--52 (1974).

\bibitem{pani-etal:15a}
P.~Pani, L.~Gualtieri, A.~Maselli, and V.~Ferrari, {\em Tidal deformations of a
  spinning compact object\/}, Phys. Rev. D {\bf 92}, 024010 (2015),
  arXiv:1503.07365.

\bibitem{letiec-casals:21}
A.~Le~Tiec and M.~Casals, {\em Spinning Black Holes Fall in Love\/}, Phys. Rev.
  Lett. {\bf 126}, 131102 (2021).

\bibitem{letiec-casals-franzin:21}
A.~Le~Tiec, M.~Casals, and E.~Franzin, {\em Tidal Love numbers of Kerr black
  holes\/}, Phys. Rev. D {\bf 103}, 084021 (2021).

\bibitem{poisson:21a}
E.~Poisson, {\em Compact body in a tidal environment: New types of relativistic
  Love numbers, and a post-Newtonian operational definition for tidally induced
  multipole moments\/}, Phys. Rev. D {\bf 103}, 064023 (2021).

\bibitem{damour-nagar:10}
T.~Damour and A.~Nagar, {\em Effective one body description of tidal effects in
  inspiralling compact binaries\/}, Phys. Rev. D {\bf 81}, 084016 (2010).

\bibitem{bini-damour-faye:12}
D.~Bini, T.~Damour, and G.~Faye, {\em Effective action approach to higher-order
  relativistic tidal interactions in binary systems and their effective one
  body description\/}, Phys. Rev. D {\bf 85}, 124034 (2012), arXiv:1202.3565.

\bibitem{vines-flanagan:13}
J.~E. Vines and E.~E. Flanagan, {\em First-post-Newtonian quadrupole tidal
  interactions in binary systems\/}, Phys. Rev. D {\bf 88}, 024046 (2013),
  arXiv:1009.4919.

\bibitem{henry-faye-blanchet:20a}
Q.~Henry, G.~Faye, and L.~Blanchet, {\em Tidal effects in the equations of
  motion of compact binary systems to next-to-next-to-leading post-Newtonian
  order\/}, Phys. Rev. D {\bf 101}, 064047 (2020).

\bibitem{manko:07}
V.~S. Manko, {\em Double-Reissner-Nordstr\"om solution and the interaction
  force between two spherical charged masses in general relativity\/}, Phys.
  Rev. D {\bf 76}, 124032 (2007).

\bibitem{manko-ruiz-sanchezmondragon:09}
V.~S. Manko, E.~Ruiz, and J.~S\'anchez-Mondrag\'on, {\em Analogs of the
  double-Reissner-Nordstr\"om solution in magnetostatics and dilaton gravity:
  Mathematical description and basic physical properties\/}, Phys. Rev. D {\bf
  79}, 084024 (2009).

\bibitem{bini-geralico-ruffini:07}
D.~Bini, A.~Geralico, and R.~Ruffini, {\em Charged massive particle at rest in
  the field of a Reissner-Nordstr\"om black hole\/}, Phys. Rev. D {\bf 75},
  044012 (2007), arXiv:gr-qc/0609041.

\bibitem{alekseev-belinski:07}
G.~A. Alekseev and V.~A. Belinski, {\em Equilibrium configurations of two
  charged masses in general relativity\/}, Phys. Rev. D {\bf 76}, 021501
  (2007).

\bibitem{landau-lifshitz:b2}
L.~D. Landau and E.~M. Lifshitz, {\em The Classical Theory of Fields, Fourth
  Edition\/} (Butterworth-Heinemann, Oxford, England, 2000).

\bibitem{lahaye-poisson:21}
M.~LaHaye and E.~Poisson, {\em Particle hanging on a string near a
  Schwarzschild black hole\/}, Phys. Rev. D {\bf 104}, 044016 (2021).

\end{thebibliography}
\end{document}